\documentclass[journal]{IEEEtran}

\usepackage{graphicx}

\ifCLASSINFOpdf
\else
\fi

\usepackage{amsmath}
\usepackage{array}

\usepackage{cite}  
\usepackage{amsmath,amssymb,amsfonts}
\usepackage{algorithmic}
\usepackage{graphicx}
\usepackage{textcomp}
\usepackage{xcolor}

\usepackage{tabularx}
\usepackage{booktabs}
\usepackage{comment}

\newlength\myindent
\usepackage{float}
\usepackage{placeins}

\usepackage{caption,setspace}

\usepackage{subcaption}
\usepackage[normalem]{ulem}  

\begin{document}

\title{Addressing intra-area oscillations and frequency stability after DC segmentation of a large AC power system}

\author{Mathieu~Robin,
        Javier~Renedo, \IEEEmembership{Senior Member, IEEE},
        Juan Carlos~Gonzalez-Torres,
        Aurelio~Garcia-Cerrada, \IEEEmembership{Senior Member, IEEE},
        Luis Rouco, \IEEEmembership{Senior Member, IEEE},
        Abdelkrim~Benchaib,  
        and~Pablo~Garcia-Gonzalez  
\thanks{This is an unabridged draft of the paper submitted on 14th June 2024 to the Journal of Modern Power Systems and Clean Energy and currently under review

This work is supported by the French Government under the program Investissements d’Avenir (ANEITE-002-01).}
}

\markboth{Preprint}%
{Shell \MakeLowercase{\textit{et al.}}: Bare Demo of IEEEtran.cls for IEEE Journals}

\maketitle

\begin{abstract}\label{sec:asbtract}
In the last decades, various events have shown that electromechanical oscillations are a major concern for large interconnected  Alternating Current (AC) power systems.  Segmentation of AC power systems with High Voltage Direct Current (HVDC) systems (DC segmentation, for short) is  a method that consists in turning large AC grids into a set of asynchronous AC clusters linked by  HVDC  links. It is a promising solution to mitigate  electromechanical oscillations  and other issues. In particular, an appropriately placed DC segmentation can stop a selected inter-area electromechanical oscillation mode. However, without supplementary controllers,  DC segmentation  will not contribute to the damping of the intra-area oscillation modes  in the remaining AC clusters  and will deteriorate the frequency stability of the  power  system. This paper aims at filling this gap and proposes the use of  DC segmentation with HVDC systems based on Voltage Source Converters (VSC-HVDC) with supplementary controllers in the converter stations:  (a) active-power supplementary controllers for frequency support among the asynchronous AC clusters and (b) a reactive-power supplementary controllers for Power Oscillation Damping (POD-Q), in order to damp the intra-area oscillation modes.  The proposed supplementary controllers and their design will be presented, and their efficiency will be demonstrated on the Nordic 44 test system  with DC segmentation  by means of non-linear time-domain simulation and small-signal stability analysis.
\end{abstract}


%
\IEEEpeerreviewmaketitle

\section{Introduction}\label{sec:Intro}
\IEEEPARstart{A}lternating Current (AC) technology is currently the dominant technology for the transmission and distribution of electrical energy. However, the transmission capability with High Voltage Alternating Current (HVAC) is limited by some technical aspects that can be overcome by the High Voltage Direct Current (HVDC) transmission technology that has been developed in the last decades \cite{luscanVisionHVDCKey2021}. Thus, many point-to-point HVDC links are already in operation around the world, and even more are planned or under construction \cite{entso-eGridMap2023, alassiHVDCTransmissionTechnology2019}. This means that electrical grids are evolving toward hybrid HVAC/HVDC power systems with a growing share of HVDC transmission \cite{gomis-bellmuntPrinciplesOperationGrids2021}. 

In parallel with the development of HVDC links, power systems have also become more and more interconnected creating large AC synchronous electrical grids. This action sums the interconnected systems' inertia values, improves frequency stability and increases the reliability of the resulting power systems. However, with the introduction of new economical objectives, power systems are operated closer to their stability limits and the risk of severe disruptions is increasing rapidly. For example, in the last decade, various events have shown that electromechanical oscillations are now a bigger threat than ever~\cite{entso-eAnalysisCEInterArea2017}. Electromechanical oscillations are a rotor angle stability phenomenon under small perturbations in a frequency range between  0.1 and 2Hz (low-frequency oscillations)~\cite{hatziargyriouDefinitionClassificationPower2021}.

Traditionally, the most cost-effective solution to damp electromechanical oscillations in power systems has been the use of supplementary controllers attached in the different devices of the power system. The most extended solution is the implementation of Power System Stabilizers (PSS) in synchronous machines~\cite{kleinFundamentalStudyInterarea1991}. Nevertheless, supplementary Power Oscillation Damping (POD) controllers in renewable power plants~\cite{Dominguez-Garcia2012,Shah2013}, Energy Storage Systems (ESS)~\cite{Jankovic2023}, Flexible Alternating Current Transmission Systems (FACTS)~\cite{roucoCoordinatedDesignMultiple2001,Mithulanathan2003}, HVDC systems based on Line Commutated Converters (LCC-HVDC)~\cite{Pierre2017} and VSC-HVDC~\cite{erikssonNewControlStructure2016,elizondoInterareaOscillationDamping2018,renedoCoordinatedDesignSupplementary2021, gonzalez-torresPowerSystemStability2019,Marinescu2021,dongAdaptivePowerOscillation2023}  systems have also proved to be effective solutions.  

However, in stressed power systems electromechanical oscillations with low damping ratio can arise in certain scenarios, even if dedicated supplementary controllers are attached in the different devices of the power system. In these situations, corrective actions may need to be taken, such as reduction of the power exchange between the different areas, for example~\cite{entso-eAnalysisCEInterArea2017}. In extremely severe cases, other corrective actions such as intentional islanding have been proposed. Intentional islanding will disconnect selected lines to split a grid into a set of stable areas – called islands – that are reconnected once the fault and its propagation have been properly addressed~\cite{liControlledPartitioningPower2010,hassaniahangarReviewIntentionalControlled2020,youSlowCoherencyBasedIslanding2004,kaisunSplittingStrategiesIslanding2003}.

Intentional islanding is only an operational action. However, recent work has analysed segmentation of AC power systems with VSC-HVDC links (DC segmentation, for short) to mitigate electromechanical oscillations \cite{robinAlgorithmDCSegmentation2023}, which is a potential solution in stressed power systems.  DC segmentation was first proposed in~\cite{clarkApplicationSegmentationGrid2008} and consists in segmenting a large AC asynchronous grid into a set of AC asynchronous grids linked by HVDC links. DC segmentation is a promising application of VSC-HVDC to tackle severe instability problems or to improve the operation of the system \cite{entso-eHVDCLinksSystem2019}. For example, a first DC segmentation project was carried out in China in 2016~ \cite{fairleyWhySouthernChina2016} to prevent potential overloading of the inter-area AC lines in case of contingency. There are different applications of DC segmentation. For example, it has the capacity to limit the propagation of disturbances \cite{clarkSofteningBlowDisturbance2008, fangBTBDCLink2009,shamiAnalysisDifferentPower2015}. Therefore, it could limit the risk of cascading failures and blackouts \cite{clarkApplicationSegmentationGrid2008}. This effect of DC segmentation has been confirmed in \cite{mousaviAssessmentHVDCGrid2013,gomilaAnalysisBlackoutRisk2023}. Additionally, DC segmentation can also improve the power exchange capacity between different areas~\cite{clarkApplicationSegmentationGrid2008,stanojevBenefitAnalysisHybrid2019, hartTransformationACTransmission2021}.

Regarding electromechanical oscillations, the work in~\cite{robinAlgorithmDCSegmentation2023} proposed an algorithm to obtain a DC-segmentation architecture for a given AC power system, targeted to mitigate electromechanical oscillations. The work in~\cite{robinDCSegmentationPromising2021} presented a comprehensive analysis of the impact of DC segmentation on transient stability, electromechanical-oscillation damping and frequency stability in power systems. It showed that DC segmentation, where VSC-HVDC systems are controlled with constant power set points, could improve transient stability and the damping of electromechanical oscillations (the DC segments act as firewalls), while the overall frequency stability is jeopardised, because each of the resulting asynchronous AC clusters have lower amounts of inertia and primary frequency support. Since in~\cite{robinAlgorithmDCSegmentation2023} the VSC-HVDC links used for DC segmentation where controlled with constant active- (P) and reactive-power (Q) injections, there is still room for further improvements.

Hence, given a power system with DC segmentation targeted to damp electromechanical oscillations (as in~\cite{robinAlgorithmDCSegmentation2023}), the target inter-area mode is eliminated. However, the following questions remain open:
\begin{itemize}
    \item Could the damping ratio of the intra-area electromechanical modes of the remaining AC clusters be increased?
    \item Could the overall frequency stability of the resulting DC-segmented system be improved?
\end{itemize}

Along these lines, the contributions of this paper are as follows:
\begin{itemize}
    \item Investigation of the use of supplementary reactive-power controllers in the VSC-HVDC systems of the DC-segmented system to damp intra-area oscillations in the remaining clusters (POD-Q controllers, for short). Two POD-Q controllers are analysed: (a) one using local frequency as input signal (based on previous proposals) and (b) another one using the frequency of the Centre Of Inertia (COI) of each AC cluster, which is a new proposal of this work.
    \item Investigation of the use of supplementary active-power/frequency controllers (FC) in the VSC-HVDC systems of the DC-segmented system to improve frequency stability.
    \item Proposal of the concept of numerical eigenvalue sensitivity, which is used in the design of POD-Q controllers. This concept is an approximation of the theoretical eigenvalue sensitivity concept and it can facilitate its implementation.
\end{itemize}
Notice that, although POD controllers \cite{elizondoInterareaOscillationDamping2018} and frequency controllers~\cite{castroProvisionFrequencyRegulation2016,zhangPrimaryFrequencySupport2021,Shinoda2023} in VSC-HVDC systems have been analysed in previous work, their application to a DC-segmented system has not received enough attention in previous work. This is essential to fully exploit DC-segmentation concept for the particular application for the mitigation of electromechanical oscillations, which is the main objective of this paper.

\section{DC segmentation of power systems and VSC-HVDC systems}\label{sec:DCs}
\begin{figure}[htb!]
\vspace{-0.3cm}
\centering
\includegraphics[width=0.65\columnwidth]{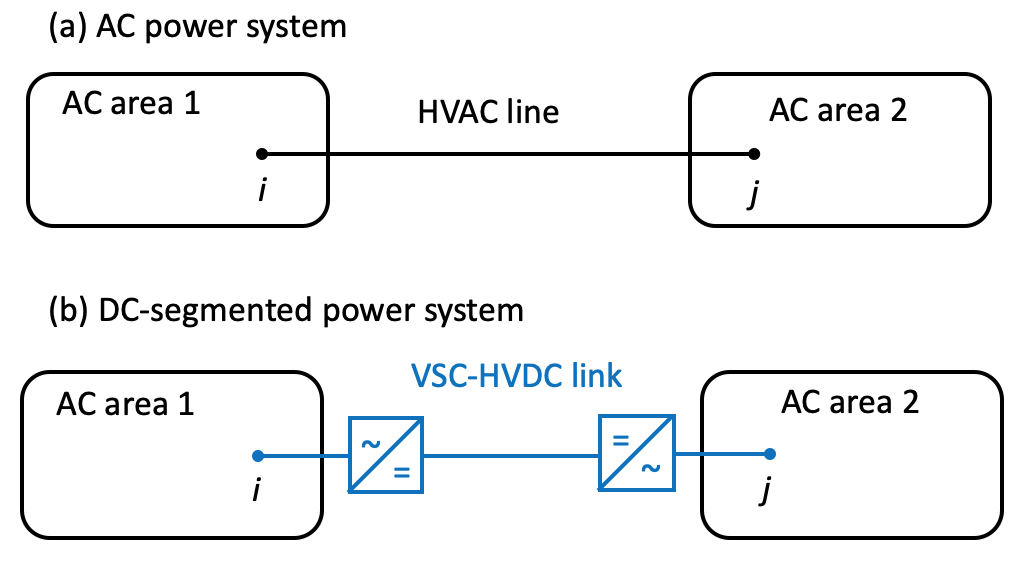}
\caption {(a) AC power system and (b) DC-segmented power system.}
\label{fig:DC_segmentation_scheme}
\end{figure}

\noindent Fig.~\ref{fig:DC_segmentation_scheme}-a shows a schematic diagram of an AC power system and Fig.~\ref{fig:DC_segmentation_scheme}-b shows the power system after DC segmentation, where an AC line has been replaced by a VSC-HVDC link, decoupling the two remaining AC systems. Notice that, in general, a DC-segmented system could have more than one VSC-HVDC link decoupling the asynchronous AC clusters of the system. This paper focuses on the application of DC segmentation for the mitigation of electromechanical oscillations. Along this line, the work in~\cite{robinAlgorithmDCSegmentation2023} proposed an algorithm to obtain a DC-segmentation architecture to mitigate the critical inter-area oscillation of the original AC power system. The algorithm used the concept of electromechanical-oscillation path proposed in \cite{chompoobutrgoolIdentificationPowerSystem2013}. The algorithm selects the AC lines to be replaced with point-to-point VSC-HVDC links, leading to a DC-segmented system with asynchronous AC clusters. This DC segmentation fully suppresses the target inter-area electromechanical mode, without impacting the local electromechanical modes of the asynchronous clusters. The work in~\cite{robinAlgorithmDCSegmentation2023} analysed DC segmentation with constant power of the VSC-HVDC links. This work goes a step further as it will consider supplementary controllers to (a) damp intra-area oscillations in the asynchronous AC clusters and (b) to provide frequency support among the asynchronous AC clusters, as it will be described in Section~\ref{sec:supplementary control}.

\begin{figure}[htb!]
\vspace{-0.3cm}
\centering
\includegraphics[width=0.85\columnwidth]{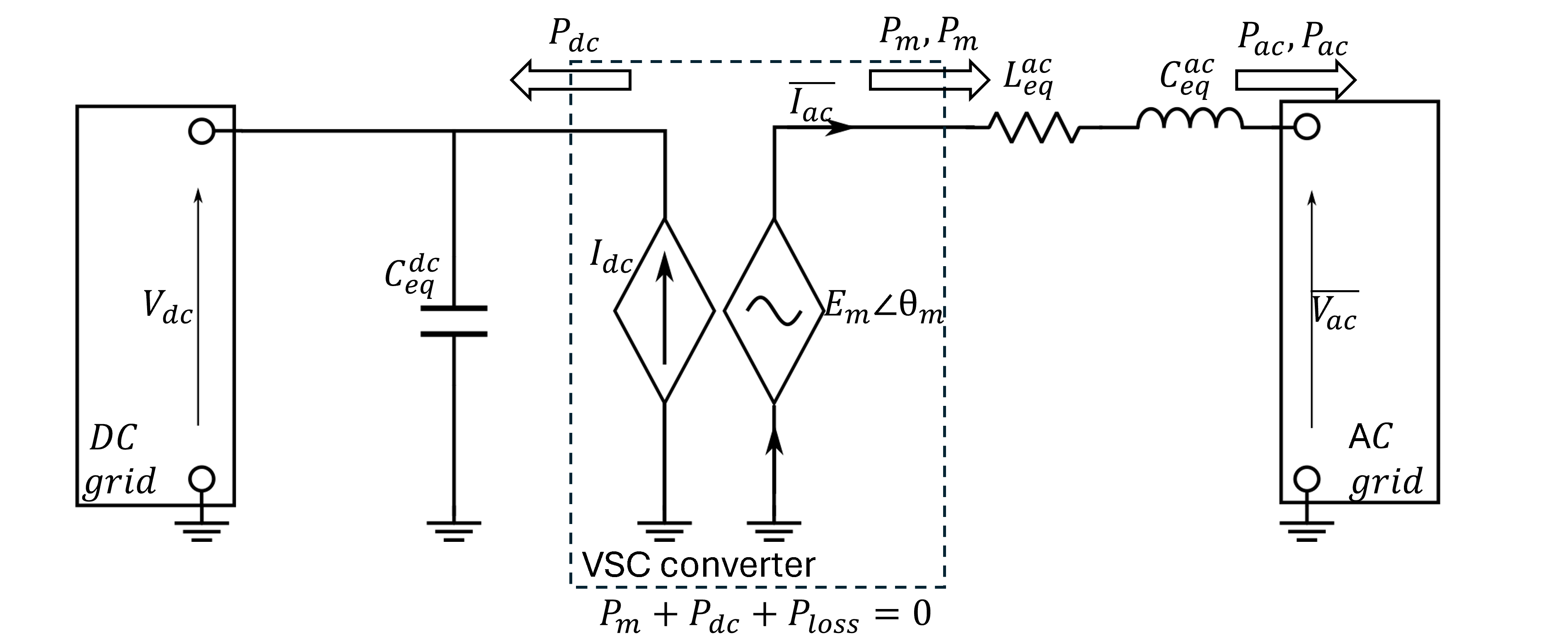}
\caption {Model of a VSC station.}
\label{fig:VSC_RMS}
\end{figure}

Electromechanical-type models of VSC-HVDC systems, also known as Root-Mean-Square (RMS) models, are used, following the guidelines of~\cite{coleGeneralizedDynamicVSC2010a}, as illustrated in Fig.~\ref{fig:VSC_RMS}. Every VSC station is modelled as a voltage source at their AC side connected by a series impedance, that aggregates the AC reactor and the transformer, and they are seen as current injections at their DC side; the AC and DC sides are coupled by the energy conservation principle. VSC stations are controlled with conventional vector control with Grid-Following (GFL) control and the model includes their operation limits, as described in~\cite{coleGeneralizedDynamicVSC2010a}. In each VSC-HVDC link, one VSC station controls its active-power injection, while the other one controls the DC voltage. Meanwhile, each VSC station controls independently its reactive power injection. Finally, the DC grid is represented with the equivalent capacitor at each DC bus and with the series resistances and inductances of the DC lines.

\FloatBarrier

\section{Supplementary controllers}\label{sec:supplementary control}
\noindent The following supplementary controllers in VSC-HVDC links of the DC-segmented case will be considered:
\begin{itemize}
    \item Frequency Control (FC): The objective is to provide frequency support between the asynchronous AC areas through the VSC-HVDC links.
    \item Power Oscillation Damping (POD) controllers for the Q injections (POD-Q): The objective is to contribute to damp intra-area oscillations in the  asynchronous AC areas by means of Q-modulation in the VSC stations of the DC segments.
\end{itemize}

\subsection{Frequency Controllers (FC)}\label{sec.FC}
\noindent In FC, the P set point of the VSC station that controls the active power through each DC segment (VSC-$i$) is given by:
\begin{equation}\label{eq:Pac_ref_FC}  
    P_{ac,i}^*=P_{ac,i}^0 + \Delta P_{ac,i}^{ref,FC}
\end{equation}
where $P_{ac,i}^0$ is a constant P set-point value, and $\Delta P_{ac,i}^{ref,FC}$ is the supplementary P set-point value provided by the FC.

\begin{figure}[htb!]
\vspace{-0.3cm}
\centering
\includegraphics[width=0.7\columnwidth]{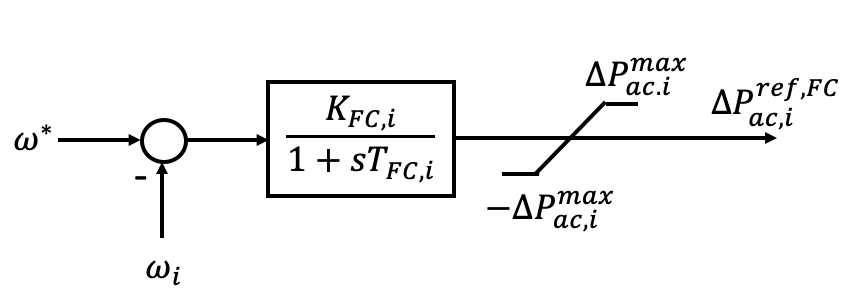}
\caption {Frequency Controller (FC) using FC-WAF.}
\label{fig:fsupport}
\end{figure}

Fig. \ref{fig:fsupport} shows the block diagram of the FC, based on strategy FC-WAF proposed in~\cite{zhangPrimaryFrequencySupport2021} for multi-terminal VSC-HVDC systems (VSC-MTDC), which uses the average frequency  of the VSC-MTDC as set point ($\omega^{*}=\bar{\omega}$) for the frequency controller. The FC uses a proportional controller with gain $K_{FC,i}$, a low-pass filter with time constant $T_{FC,i}$ and a saturation parameter $\pm \Delta P_{ac,i}^{max}$. The frequency set point for the FC is given by:
\begin{equation}\label{eq:FC_WAF_wref}  
    \omega^{*} = \bar{\omega} = \frac{\omega_i+\omega_j}{2}
\end{equation}
where $\omega_{i}$ and $\omega_{j}$ are the frequency measured at the two AC terminals of the VSC-HVDC link (in pu).


Notice that in point-to-point VSC-HVDC links, strategy FC-WAF~\cite{zhangPrimaryFrequencySupport2021} is equivalent a FC controller proportional to the frequency difference between the frequencies of each terminal of the link as long as $K_{FC,i}'=K_{FC,i}/2$:
\begin{eqnarray}\label{eq:FC}  
    \Delta P_{ac,i}^{ref,FC} &=& \frac{K_{FC,i}}{1+sT_{FC,i}} (\bar{\omega}-\omega_i) \\ \nonumber
    &=& \frac{K_{FC,i}}{1+sT_{FC,i}} \big[ \frac{\omega_i+\omega_j}{2}-\omega_i \big] \\ \nonumber
    &=& \frac{K_{FC,i}/2}{1+sT_{FC,i}} (\omega_j-\omega_i) = \frac{K_{FC,i}'}{1+sT_{FC,i}} (\omega_j-\omega_i). 
\end{eqnarray}


With the FC of Fig.~\ref{fig:fsupport}, frequency support among the asynchronous AC areas through the VSC-HVDC links is achieved, independently of the location of the disturbance, since the FC uses frequency measurements at both ends of the VSC-HVDC links. 

\subsection{POD-Q controllers}\label{sec.POD_Q}
\noindent Two POD controllers for the reactive-power injections of the VSC stations (POD-Q) will be analysed and compared in this work:
\begin{itemize}
    \item POD-Q-LF: It modulates Q injections of the VSC and it using local measurement of the frequency at the AC connection point.
    \item POD-Q-FCOI: It modulates Q injection of the VSC using global measurements of the frequency of the Centre Of Inertia (COI) of the AC area at which the VSC is connected to. 
\end{itemize}
The Q set point of each VSC-$i$ is given by:
\begin{equation}\label{eq:POD_Q}  
    Q_{ac,i}^*=Q_{ac,i}^0 + \Delta Q_{ac,i}^{ref,POD}
\end{equation}
where $Q_{ac,i}^0$ is a constant Q set-point value, and $\Delta Q_{ac,i}^{ref,POD}$ is the supplementary Q set-point value provided by POD-Q controller.

\begin{figure}[htb!]
\vspace{-0.3cm}
\centering
\includegraphics[width=1.0\columnwidth]{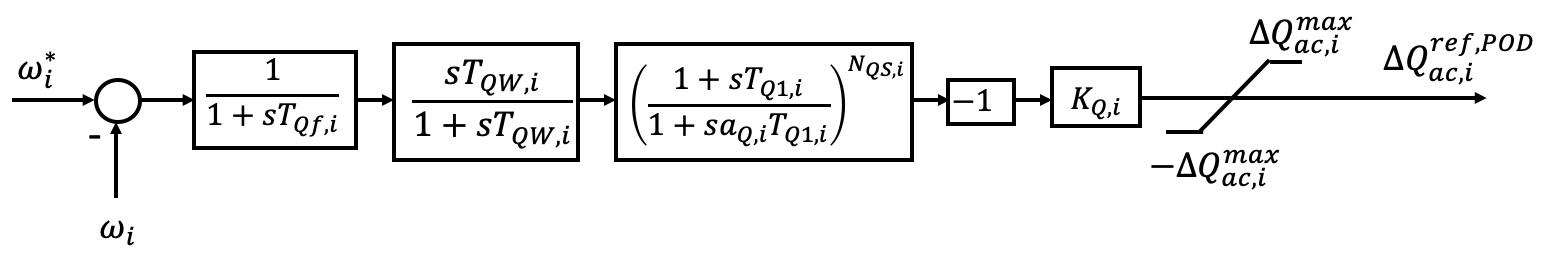}
\caption {POD-Q controller.}
\label{fig:PODQ}
\end{figure}

The two POD-Q controllers above can be described with the scheme of Fig.~\ref{fig:PODQ}. The input signal of the controller is the frequency error (in pu) between the frequency set point ($\omega_{i}^*$) and the frequency measured at the AC connection point of the VSC station ($\omega_i$). The POD-Q controller has a low-pass filter (with time constant $T_{Qf,i}$), a wash-out filter (with time constant $T_{QW,i}$), a lead/lag filter (with time constants $T_{Q1,i}$, $T_{Q2,i}=a_{Q,i}T_{Q1,i}$, filtering ratio $a_{Q,i}$ and $N_{QS,i}$ lead/lag networks), a controller gain $K_{Q,i}$ and a saturation parameter $\pm  \Delta Q_{ac,i}^{max}$. 

\subsubsection{POD-Q-LF}\label{sec.POD_Q_LF} $ $ \\
In POD-Q-LF controller, the frequency set point of Fig.~\ref{fig:PODQ} is the nominal frequency in pu: $\omega_{i}^{*}=\omega_{0}=1$~pu. Hence, it uses local measurements. This scheme of POD-Q controller is based on previous schemes proposed in the literature~\cite{roucoDampingElectromechanicalOscillations1996,Jankovic2023}.

\subsubsection{POD-Q-FCOI}\label{sec.POD_Q_FCOI} $ $ \\
In POD-Q-FCOI controller, proposed in this work, the frequency set point of Fig.~\ref{fig:PODQ} is calculated as the frequency of the COI (in pu) of the AC area which station VSC-$i$ is connected:
\begin{equation}\label{eq:f_COI}  
    \omega_{i}^{*}= \omega_{COI,i}= \sum_{k \in A_i}  \frac{H_k}{H_T} \omega_k, \mbox{ } H_T=\sum_{j \in A_i} H_j.
\end{equation}
This strategy was motivated by POD-Q-WAF controller proposed in~\cite{renedoCoordinatedDesignSupplementary2021} for VSC-MTDC systems embedded in an AC system, which used as frequency set point the weighted-average frequency of the VSC-MTDC system. However, the application and the controller proposed here are different, since the VSC-HVDC are interconnecting asynchronous AC areas, and the frequency of the COI of each AC area are used as frequency set points for POD-Q controllers. In POD-Q-FCOI controller, each VSC compares the frequency set point with $\omega_{i}^{*}= \omega_{COI,i}$ with its own frequency ($\omega_i$). Notice that the usefulness of the speed of the COI for power-system-stability-tailored controllers was illustrated in~\cite{LuisDM2019}.

\vspace{-0.5cm}
\subsection{Design of POD-Q controllers}\label{sec.POD_Q_design}
\noindent The method of the design of POD-Q controllers is based on the work in~\cite{roucoCoordinatedDesignMultiple2001}, which uses the concept of eigenvalue sensitivities~\cite{pagolaSensitivitiesResiduesParticipations1989}. The work in~\cite{renedoCoordinatedDesignSupplementary2021} used this approach for the design of POD controllers in VSC-HVDC systems. As a novelty of this work, the concept of \emph{numerical eigenvalue sensitivity} is proposed (instead of the theoretical eigenvalue sensitivity), which can facilitate the design of POD controllers in cases with limited information of the state matrices of the linearised system. 

It is assumed that a linearised model of the power system is available. A POD-Q controller in VSC-$j$ will be used to damp a target electromechanical mode $i$:
\begin{equation}
 \label{eq:lambda0}  
 \lambda_i^0 = \sigma_i^0 \pm j \omega_i^0
\end{equation}
The damping ratio of the target electromechanical mode of the original system is called $\zeta_i^0$ and the required damping ratio is called $\zeta_i^d$. Then, the estimated target electromechanical mode can be approximated as:
\begin{equation}
 \label{eq:lambdad}  
 \lambda_i^d = - \zeta_i^d \omega_i^0 \pm j \omega_i^0
\end{equation}
The sensitivity of electromechanical mode $i$ to changes in the gain of POD-Q controller $j$, $K_{Q,j}$, is defined as~\cite{pagolaSensitivitiesResiduesParticipations1989}:
\begin{equation}
 \label{eq:Sij}  
 S_{ij} = \frac{\partial \lambda_i}{\partial K_{Q,j}} 
\end{equation}

Similarly, the non-compensated sensitivity of mode $i$ to changes in the gain of the non-compensated POD-Q controller $j$ ($S_{ij}^{NC}$) is defined as:
\begin{eqnarray}
 \label{eq:SijNC}  
 S_{ij}^{NC} &=& S_{ij}(T_{Q1,j}=0) = \frac{\partial \lambda_i}{\partial K_{Q,j}}\bigg|_{T_{Q1,j}=0} \\ \nonumber
 &=& |S_{ij}^{NC}| \angle \varphi_{ij}^{NC} 
\end{eqnarray} 
The \emph{numerical non-compensated eigenvalue sensitivity}, proposed in this work,
can be calculated as:
\begin{equation}
 \label{eq:hatSijNC}  
 \hat S_{ij}^{NC} = \frac{\lambda_i^{NC}-\lambda_i^0}{\Delta K_{Q,j}}
\end{equation}
where $\lambda_i^{NC}$ is the non-compensated eigenvalue, i.e., the new eigenvalue with the POD-Q implemented without its lead/lag filter ($T_{Q1,j=0}$) and with a small gain $\Delta K_{Q,j}$. Naturally, $\hat S_{ij}^{NC} \approx S_{ij}^{NC}$.

\begin{figure}[htb!]
\vspace{-0.3cm}
\centering
\includegraphics[width=0.3\textwidth]{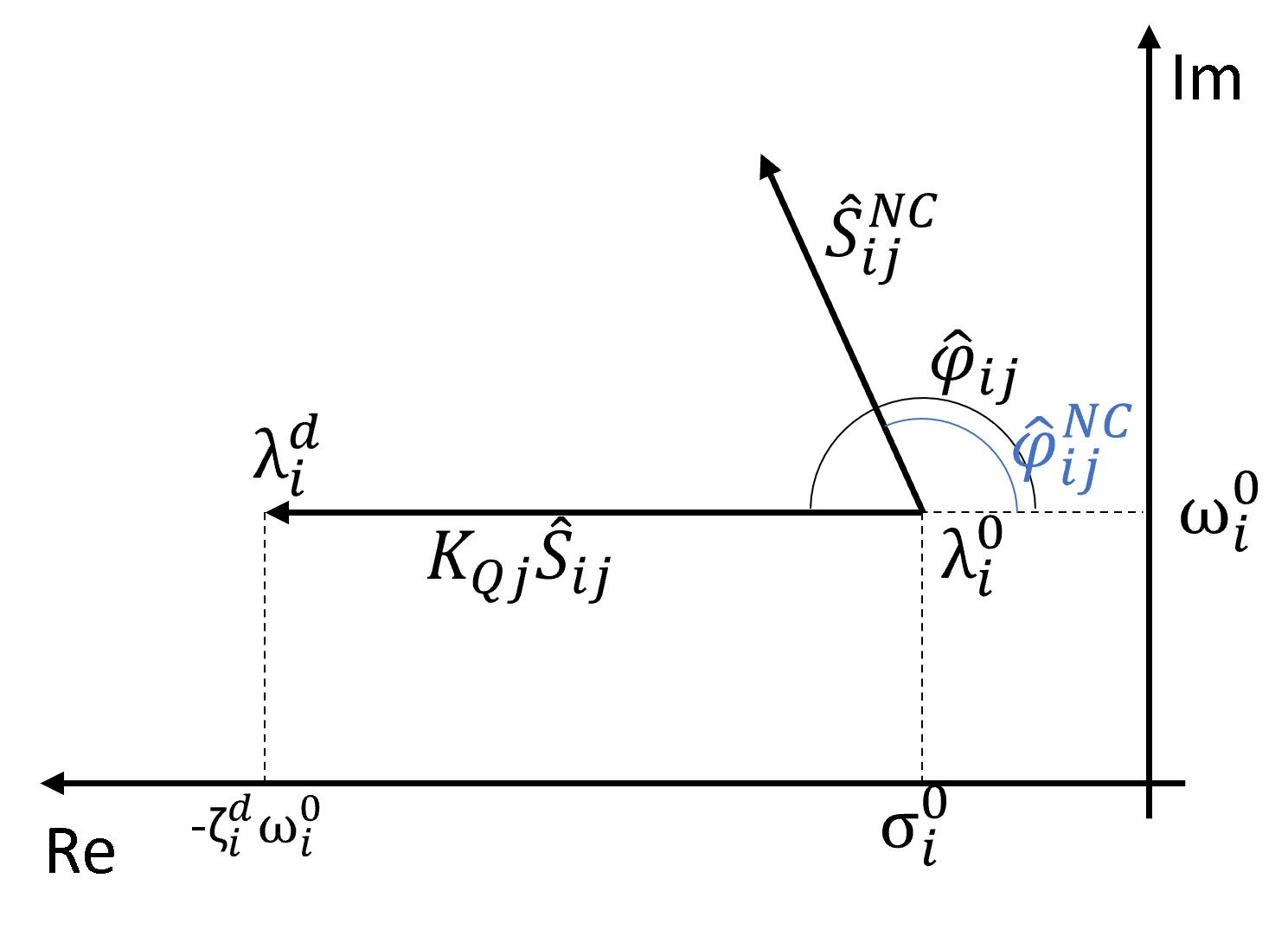}
\caption {Geometric interpretation of eigenvalue sensitivities.}
\label{fig:sensitivities}
\end{figure}

Fig.~\ref{fig:sensitivities} shows the geometric interpretation of eigenvalue sensitivities~\cite{roucoCoordinatedDesignMultiple2001,renedoCoordinatedDesignSupplementary2021}. An effective POD-Q will move the target electromechanical mode to the left-hand side of the complex plane, as illustrated in Fig.~\ref{fig:sensitivities}. The lead/lag filter is used to ensure that the eigenvalue sensitivity has phase as close to 180° as possible, while gain $K_{Q,j}$ (see Fig.~\ref{fig:PODQ}) is used to move the eigenvalues further to the left-hand side of the complex plane and to achieve the required damping ratio.

The design of the POD-Q controller of each VSC-$j$ consists of two steps~\cite{roucoCoordinatedDesignMultiple2001}:
\begin{enumerate}
    \item {\em Design of the lead/lag filter:} Its objective is to obtain a phase of the eigenvalue sensitivity ($\hat S_{ij}$) as close to $180^{{\rm o}}$ as possible~\cite{roucoCoordinatedDesignMultiple2001}:

    ${\rm phase\;lead } \mbox{\space}(\hat{\varphi}_{ij}^{NC} \geq 0 $): 
    \begin{equation}
        a_{Q,j} = \frac{1 - \sin \phi_{ij}}{1 + \sin \phi_{ij}} \le 1,  \;\phi_{ij} = \frac{\pi-\hat{\varphi}_{ij}^{NC}}{N_{QS,j}}. \label{eq.phaselead1}
    \end{equation}
    
    ${\rm phase\;lag } \mbox{\space} (\hat{\varphi}_{ij}^{NC} < 0 $): 
    \begin{equation}
        a_{Q,j} = \frac{1 + \sin \phi_{ij}}{1 - \sin \phi_{ij}} >1, \; \phi_{ij} = \frac{\pi+\hat{\varphi}_{ij}^{NC}}{N_{QS,j}}. \label{eq.phaselag2}
    \end{equation}

    with $\hat{\varphi}_{ij}^{NC}=\angle{(\hat S_{ij}^{NC})}$. 
    
    Finally, the  time constants of the lead/lag filter ($T_{Q1,j}$ and $T_{Q2,j}$) are selected to achieve maximum lead/lag phase compensation at the frequency of the target electromechanical mode~\cite{Larsen_PSS_part2_1981}:
    
\begin{equation}\label{eq:leadlag_TQ1}
    T_{Q1,j} = \frac{1}{\omega_i^0 \sqrt{a_{Q,j}}}, \mbox{ and \space } T_{Q2,j}=a_{Q,j}T_{Q1,j}.
\end{equation}

The resulting  numerical eigenvalue sensitivity can be then calculated as:
\begin{equation}
 \label{eq:hatSij_final}  
 \hat S_{ij} = \hat{S}_{ij}^{NC} \Big[\frac{1+s T_{Q1,j}}{1+s a_{Q,j}T_{Q1,j}} \Big]_{s=\lambda_i^0}^{N_{QS,j}} = |\hat{S}_{ij}^{NC}| \angle{\hat{\varphi}_{ij}}
\end{equation}

    \item {\em Calculation of the controller gain:} Gain $K_{Q,j}$ is calculated to obtain the required damping ratio of the target electromechanical mode: 
\begin{equation}
 \label{eq:KSj_lambda_d}  
 \lambda_i^d \approx \lambda_i^0 + K_{Q,j}\hat S_{ij}.  
\end{equation}
    \begin{equation}
 \label{eq:KSj}  
 K_{Q,j} = \gamma_j \frac{|\lambda_i^d-\lambda_i^0|}{|\hat S_{ij}|}
\end{equation}
with $K_{Q,j} \in [-K_{Q,j}^{max}, K_{Q,j}^{max}]$ and $\gamma_j=1$ if a positive gain is required and $\gamma_j=-1$ if a negative gain is required.
\end{enumerate}

\section{Case study and results}\label{sec:results}
\noindent Nordic 44 test system~\cite{jakobsenNordic44Test2018} is considered to analyse DC segmentation and the supplementary controllers described in Section~\ref{sec:supplementary control}.  Nordic 44 test system was initially implemented within iTesla project as an application example of the OpenIPSL library, implemented in the Modelica language~\cite{vanfrettiITeslaPowerSystems2016, baudetteOpenIPSLOpenInstancePower2018}. The version used in this paper is the one updated by the ALSETlab. The simulations will be carried out using the Dymola environment. OpenIPSL can be used for non-linear electromechanical time-domain simulation, and small-signal stability analysis. The information about the scenario considered is provided in Section~\ref{sec:Appendix-N44_AC_base_case} of the Appendix.

\begin{figure*}[htb!]
\vspace{-0.3cm}
\centering
\includegraphics[width=0.85\textwidth]{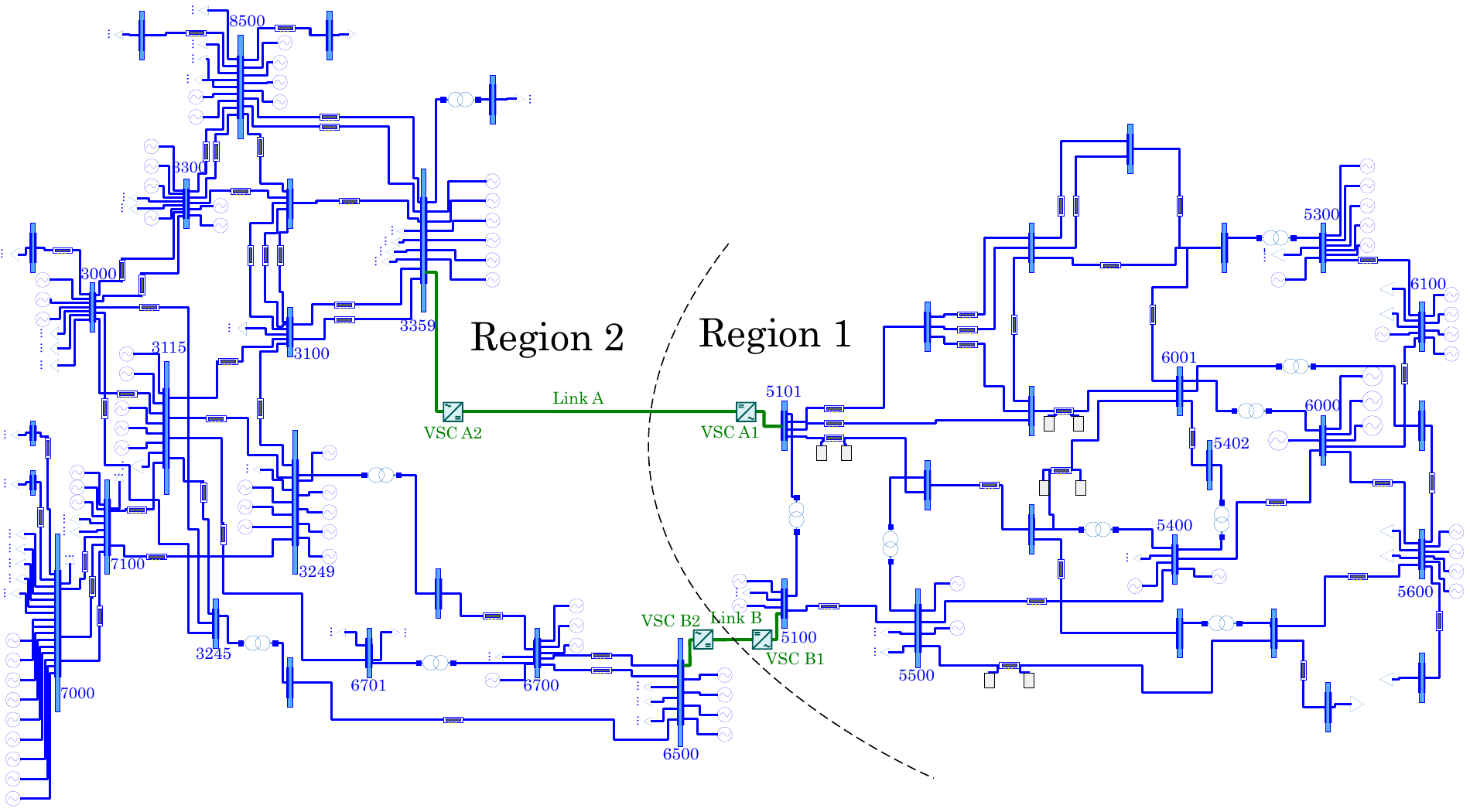}
\caption {DC-segmented N44 test system under Dymola.}
\label{fig:N44system}
\end{figure*}

Initially, Nordic 44 test system is an AC power system. The starting point for the work presented in this paper is the architecture for DC-segmentation obtained by the algorithm for the mitigation of electromechanical oscillations proposed in~\cite{robinAlgorithmDCSegmentation2023}, as shown in Fig.~\ref{fig:N44system}. VSC-HVDC link A has a rating of 3500 MVA, while VSC-HVDC link B has a rating of 800 MVA. Supplementary controllers for frequency support among the two AC areas (FC) and power-oscillation damping for the intra-area modes by means of Q-modulation at the VSC stations (POD-Q controllers) will be analysed (see Section~\ref{sec:supplementary control}).  

Four cases will be analysed and compared: 
\begin{itemize}
    \item \textbf{AC base case}: The initial Nordic 44 test system. It corresponds to the case of Fig.~\ref{fig:N44system} with AC lines instead of the DC links.
    \item \textbf{DCs constant PQ}: The DC-segmented case without supplementary controllers. 
    \item \textbf{DCs f-support, POD-Q-LF}: The DC-segmented case with frequency support (FC) and POD-Q using local frequency (POD-Q-LF).  
    \item \textbf{DCs f-support, POD-Q-FCOI}: The DC-segmented case with frequency support (FC) and POD-Q using the frequency of the centre of inertia of each AC area (POD-Q-FCOI).  
\end{itemize}

In the operating point, the power flows through the VSC-HVDC links is the same as the power flows through the AC lines of the base case: 447~MW and 1548~MW (from Region 1 to Region 2). The four cases have been compared by means of small-signal stability analysis (see Section~\ref{sec:SSA}) and non-linear time-domain simulations (see Section~\ref{sec:simu}).

\subsection{Design of POD-Q controllers}
\noindent Table~\ref{tab:eigvbasecase} presents the main modes of the Nordic 44 test system in the AC case and in the DC-segmented case without supplementary controllers. All the electromechanical modes with a damping lower than 20\% in at least one of the cases have been included. The table also indicates the location of each mode. Results show that the critical inter-area mode (mode 1), which has a low damping ratio, has been mitigated with the DC-segmentation architecture obtained with the algorithm proposed in~\cite{robinAlgorithmDCSegmentation2023}. Nevertheless, there are still intra-area modes of each asynchronous AC area. POD-Q controllers (POD-Q-LF and POD-Q-FCOI described in Section~\ref{sec.POD_Q}) will be used at every VSC station to increase the damping ratio of some intra-area modes. POD-Q controllers will be designed following the methodology presented in Section~\ref{sec.POD_Q_design}.

\begin{table}[htb!]
\centering
\caption{Poorly-damped  electromechanical modes of the two  initial cases.}
\label{tab:eigvbasecase} 
\scalebox{0.65}{
\begin{tabular}{@{}c|cc|cc|c@{}}
\toprule
    &   \multicolumn{2}{c}{AC base case} &  \multicolumn{2}{c}{DCs constant PQ} & Region of\\
N0. &   $\zeta$ (\%) & Freq. (Hz)&  $\zeta$ (\%) & Freq. (Hz) & the mode\\ \midrule
1	&	1.85	&	0.39	&	-	&	-	&	Inter-area	\\
2	&	5.45	&	0.83	&	5.70	&	0.80	&	R1	\\
3	&	11.68	&	0.88	&	11.92	&	0.87	&	R2	\\
4	&	12.11	&	0.75	&	11.73	&	0.72	&	R2	\\
5	&	-	&	-	&	12.06	&	1.12	&	R2	\\
6	&	12.12	&	1.07	&	12.14	&	1.07	&	R1	\\
7	&	12.22	&	0.54	&	12.38	&	0.50	&	R2	\\
8	&	13.12	&	0.98	&	13.12	&	0.95	&	R2	\\
9	&	13.57	&	1.23	&	17.37	&	1.27	&	R1	\\
10	&	15.69	&	1.10	&	14.81	&	1.12	&	R1	\\
\bottomrule
\end{tabular}%
}
\end{table}

POD-Q controllers of VSCs of region 1 (ie. VSC-A1 or 5101 and VSC-B1 or 5100) will target mode 2 (with a damping ratio of $\zeta=5.7$~\% and a frequency of 0.80~Hz) while the VSC of region 2 (ie. VSC-A2 or 3359 and VSC-B2 or 6500) will target mode 4 (with a damping ratio of $\zeta=11.92$~\% and a frequency of 0.72~Hz). Each POD-Q controller will be designed independently to obtain a 15~\%  damping for the targeted mode. Gain step used for the calculation of the numerical eigenvalue sensitivity is $\Delta K_{Q,j}=20$~pu (nominal pu). Pre-defined parameters of the POD-Q controllers are: $T_{Qf,j}=0.1$~s, $T_{QW,j}=5$~s, $N_{QS,j}=2$ and $\pm \Delta Q_{ac,i}^{max}= \pm 0.1$~pu. Maximum allowed POD-Q gain is set to $K_{Q,j}^{max}= \pm 400$~pu. 

Table~\ref{tab:PODQdesign} shows the parameters obtained for POD-Q-LF (local) and POD-Q-FCOI (global) controllers obatined with the design method of Section~\ref{sec.POD_Q_design}. Notice that POD-Q gains reach their maximum allowed value. This is due to the fact that POD controllers in VSC-HVDC systems can be very effective if the VSC stations are well located to damp electromechanical oscillations~\cite{renedoCoordinatedDesignSupplementary2021}. However, POD-Q controllers in a DC-segmented system may not be in the best location to damp electromechanical oscillations. Precisely, the aim of this work is to analyse the capability of damping intra-area electromechanical modes in DC-segmented system with a pre-defined scheme. It is also worth to highlight that some POD-Q gains have opposite signs, which is related to the location of the VSC stations in each AC area.  

\begin{table}[htb!]
\centering
\caption{Parameters of the POD-Q controllers.}
\label{tab:PODQdesign}
\scalebox{0.65}{
\begin{tabular}{@{}cc|cccc@{}}
\toprule
Case	&	VSC	&	$K_{Q,j}$ (pu)	&	$T_{Q1,j}$ (s)	&	$a_{Q,j}$	\\
 \midrule
POD-Q-LF	&	A1 (5101)	&	-400	&	0.2847	&	0.486	\\
	&	B1 (5100)	&	-400	&	0.2847	&	0.486	\\
	&	A2 (3359)	&	400	&	0.2223	&	0.999	\\
	&	B2 (6500)	&	400	&	0.2975	&	0.558	\\  \midrule
POD-Q-FCOI  &	A1 (5101)	&	-400	&	0.2750	&	0.521	\\
	&	B1 (5100)	&	-400	&	0.2776	&	0.511	\\
	&	A2 (3359)	&	400	&	0.2402	&	0.856	\\
	&	B2 (6500)	&	400	&	0.2889	&	0.592	\\
\bottomrule
\end{tabular}%
}
\end{table}

\subsection{Small-signal stability analysis}\label{sec:SSA}

\noindent Table~\ref{tab:eigv} shows the main electromechanical modes of the system for the 4 cases. 

\begin{table*}[htb!]
\centering
\caption{Poorly-damped electromechanical modes of the four scenarios.}
\label{tab:eigv}
\scalebox{0.65}{
\begin{tabular}{@{}c|cc|cc|cc|cc|c@{}}
\toprule
    &   \multicolumn{2}{c}{AC base case} &  \multicolumn{2}{c}{DCs constant PQ} &   \multicolumn{2}{c}{DCs f-support POD-Q-LF}  &   \multicolumn{2}{c}{DCs f-support POD-Q-FCOI} & Region of \\
N0. &   $\zeta$ (\%) & Freq. (Hz)&  $\zeta$ (\%) & Freq. (Hz) &   $\zeta$ (\%) & Freq. (Hz)&  $\zeta$ (\%) & Freq. (Hz) & the mode \\ \midrule
1	&	1.85	&	0.39	&	-	&	-	&	-	&	-	&	-	&	-	&	Inter-area	\\
2	&	5.45	&	0.83	&	5.70	&	0.80	&	11.53	&	0.84	&	12.40	&	0.83	&	R1	\\
3	&	11.68	&	0.88	&	11.92	&	0.87	&	12.30	&	0.88	&	12.33	&	0.88	&	R2	\\
4	&	12.11	&	0.75	&	11.73	&	0.72	&	15.52	&	0.74	&	15.40	&	0.74	&	R2	\\
5	&	-	&	-	&	12.06	&	1.12	&	12.44	&	1.10	&	12.16	&	1.07	&	R2	\\
6	&	12.12	&	1.07	&	12.14	&	1.07	&	12.15	&	1.07	&	12.48	&	1.10	&	R1	\\
7	&	12.22	&	0.54	&	12.38	&	0.50	&	15.02	&	0.48	&	16.45	&	0.47	&	R2	\\
8	&	13.12	&	0.98	&	13.12	&	0.95	&	16.05	&	0.97	&	15.61	&	0.97	&	R2	\\
9	&	13.57	&	1.23	&	17.37	&	1.27	&	20.68	&	1.31	&	20.68	&	1.31	&	R1	\\
10	&	15.69	&	1.10	&	14.81	&	1.12	&	14.30	&	1.21	&	14.55	&	1.20	&	R1	\\
\bottomrule
\end{tabular}%
}
\end{table*}

Results show that:
\begin{itemize}
    \item The inter-area mode (mode 1 with a damping ratio of 1.85~\% and a frequency of 0.39~Hz) has been suppressed by the DC segmentation. 
    \item The main intra-area mode of region 1 (mode 2 with a damping ratio of 5.7~\% and a frequency of 0.8~Hz in the DC-segmented case without control) is well damped by the POD-Q controllers (+5.97~\% with POD-Q-LF and +7.04~\% with POD-Q-FCOI).
    \item The main intra-area mode of region 2 (mode 4 with a damping ratio of 11.73~\% and a frequency of 0.72HZ in the DC-segmented case without control) is well damped by the POD-Q controllers (+3.62~\% with POD-Q-LF and +3.67~\% with POD-Q-FCOI).
    \item Meanwhile, the damping ratios of the rest of the electromechanical modes do not change significantly. 
\end{itemize}

Notice that in the DC-segmented case, results obtained obtained with POD-Q-FCOI (global measurements) are slightly better than the ones obtained with POD-Q-LF (local measurements), although the improvements are similar. For this reason, taking into account that the implementation of POD-Q using local measurements is much easier, for this test system the use of POD-Q-LF seems to be a better solution than POD-Q-FCOI.

\subsection{Non-linear time-domain simulation}\label{sec:simu}
\subsubsection{Small perturbation}\label{sec:small}
$ $ \\
The disconnection of line 6001-5301  at Region 1 at $t=1$ s (a small perturbation) has been simulated. The purpose of this simulation is to analyse electromechanical oscillations.

\begin{figure} [htb!]
     \centering
     \begin{subfigure}[b]{0.37\textwidth}
         \centering
         \includegraphics[width=\textwidth,trim={0cm 0.1cm 0.7cm 0.4cm},clip]{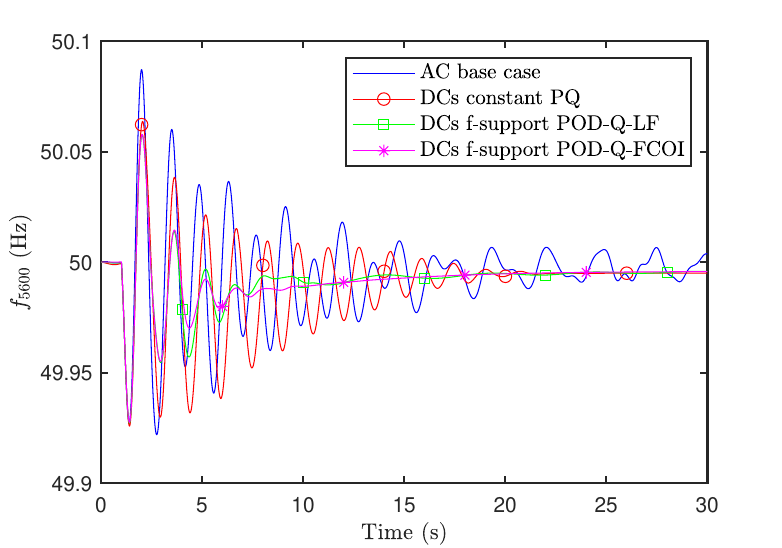}
         \caption{Region 1}
         \label{fig:small_R1_G5600}
     \end{subfigure}
     \hfill
     \begin{subfigure}[b]{0.37\textwidth}
         \centering
         \includegraphics[width=\textwidth,trim={0cm 0.1cm 0.7cm 0.4cm},clip]{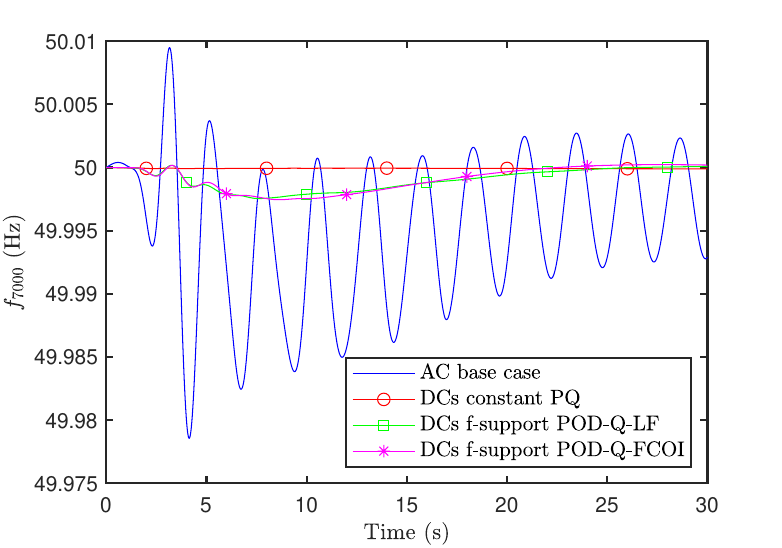}
         \caption{Region 2}
         \label{fig:small_R1_G7000}
     \end{subfigure}
     \hfill
        \caption{Frequency of two representative generators after the disconnection of line 6001-5301 (Region 1).}
        \label{fig:small_R1_G}
\end{figure}

\begin{figure} [htb!]
     \centering
     \begin{subfigure}[b]{0.37\textwidth}
         \centering
         \includegraphics[width=\textwidth,trim={0cm 0.1cm 0.7cm 0.4cm},clip]{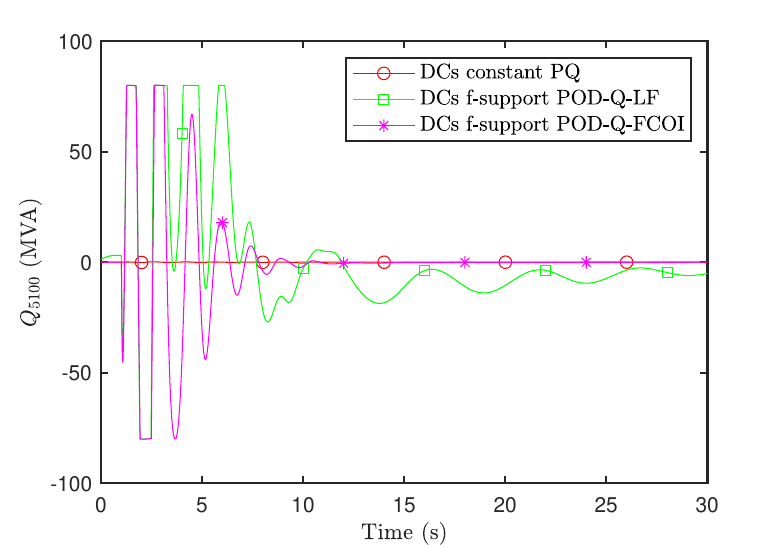}
         \caption{VSC A1 (5100)}
         \label{fig:small_R1_Q5100}
     \end{subfigure}
     \hfill
     \begin{subfigure}[b]{0.37\textwidth}
         \centering
         \includegraphics[width=\textwidth,trim={0cm 0.1cm 0.7cm 0.4cm},clip]{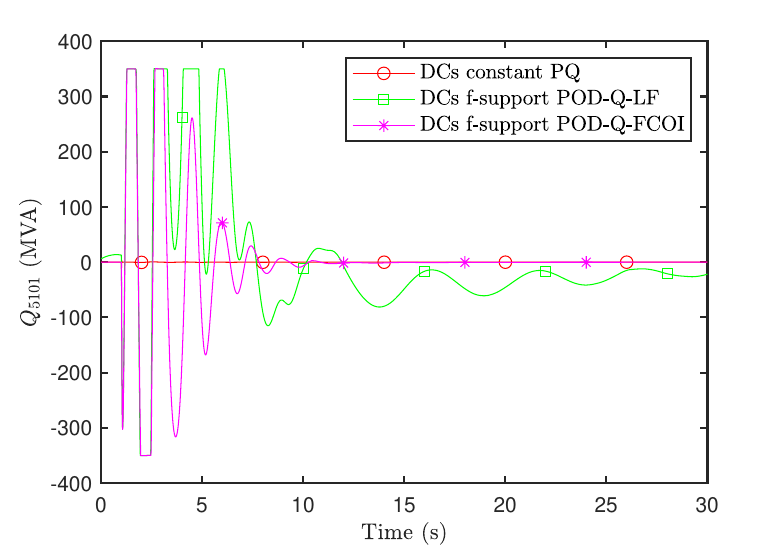}
         \caption{VSC B1 (5101)}
         \label{fig:small_R1_Q5101}
     \end{subfigure}
     \hfill
        \caption{Reactive power injection at the two VSC of Region 1 after the disconnection of line 6001-5301 (Region 1).}
        \label{fig:small_R1_Q}
\end{figure}

Fig.~\ref{fig:small_R1_G} shows the frequency of two representative generators (5600 in Region 1 and 7000 in Region 2) and Fig.~\ref{fig:small_R1_Q} shows the Q injections of the VSCs of Region 1 after the disconnection of the line in Region 1. 
An poorly-damped inter-area oscillation can be clearly observed in AC base case, which corresponds to mode 1 of Table~\ref{tab:eigv} (see Fig.~\ref{fig:small_R1_G}). Results confirm the suppression of the inter-area mode in the 3 DC-segmented cases and the limitation of the first local mode of Region 1 (mode 2 of the Table~\ref{tab:eigv}) in the two DC-segmented cases with both POD-Q-LF and POD-Q-FCOI. Furthermore, POD-Q-LF and POD-Q-FCOI controllers damp  intra-area oscillations successfully (see Fig.~\ref{fig:small_R1_G}) by means of reactive-power modulation (Fig.~\ref{fig:small_R1_Q}). Notice also that in the DC-segmented case with constant power control, the perturbation, that occurs at Region 1 is not propagated to Region 2 and, therefore, the frequency at buses in Region 2 remain constant to 50 Hz (see Fig.~\ref{fig:small_R1_G7000}). On the contrary, a small transient on the frequency at Region 2 is observed in the DC-segmented cases with supplementary controllers. This is due to the effect of the frequency controllers in the VSC-HVDC links.

\subsubsection{Disconnection of generator}\label{sec:loss_gene}
$ $ \\
The disconnection of a generator at bus 5300 (G5300-1 at Region 1) at $t=1$ s has been simulated. This generator has a nominal power of 1200 MVA and it was injecting 723 MW before the event, and it is the one with highest active-power injection before the event. The purpose of this simulation is to analyse frequency stability. \color{black}

\begin{figure} [htb!]
     \centering
     \begin{subfigure}[b]{0.37\textwidth}
         \centering
         \includegraphics[width=\textwidth,trim={0cm 0.1cm 0.7cm 0.4cm},clip]{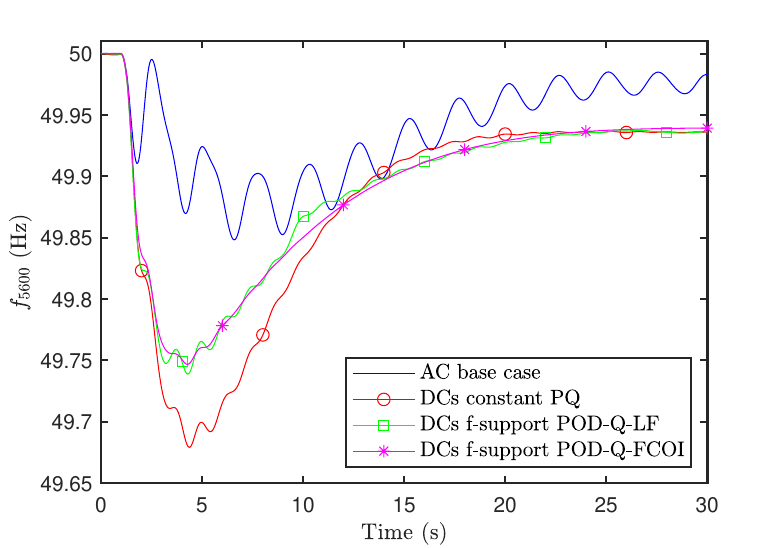}
         \caption{Region 1}
         \label{fig:loss_gene_R1_G5600}
     \end{subfigure}
     \hfill
     \begin{subfigure}[b]{0.37\textwidth}
         \centering
         \includegraphics[width=\textwidth,trim={0cm 0.1cm 0.7cm 0.4cm},clip]{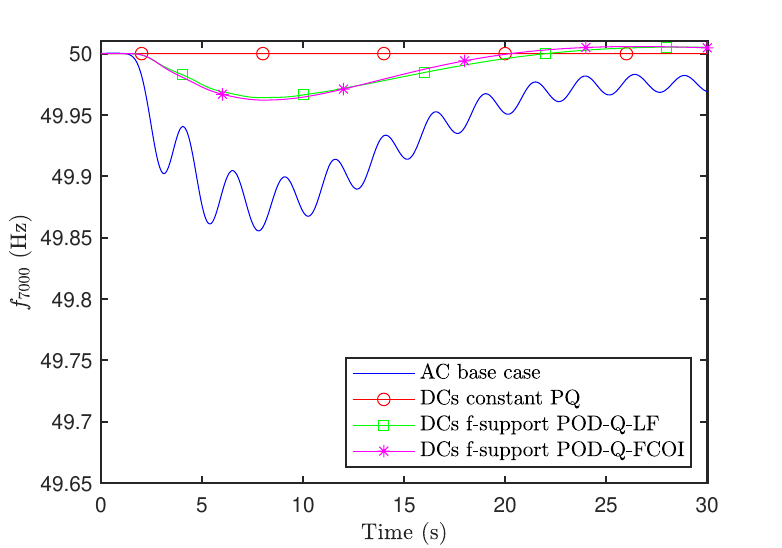}
         \caption{Region 2}
         \label{fig:loss_gene_R1_G7000}
     \end{subfigure}
     \hfill
        \caption{Frequency of two representative generators after the disconnection of a generator at bus 5300 (Region 1).}
        \label{fig:loss_gene_R1_G}
\end{figure}

\begin{figure} [htb!]
     \centering
     \begin{subfigure}[b]{0.37\textwidth}
         \centering
         \includegraphics[width=\textwidth,trim={0cm 0.1cm 0.7cm 0.4cm},clip]{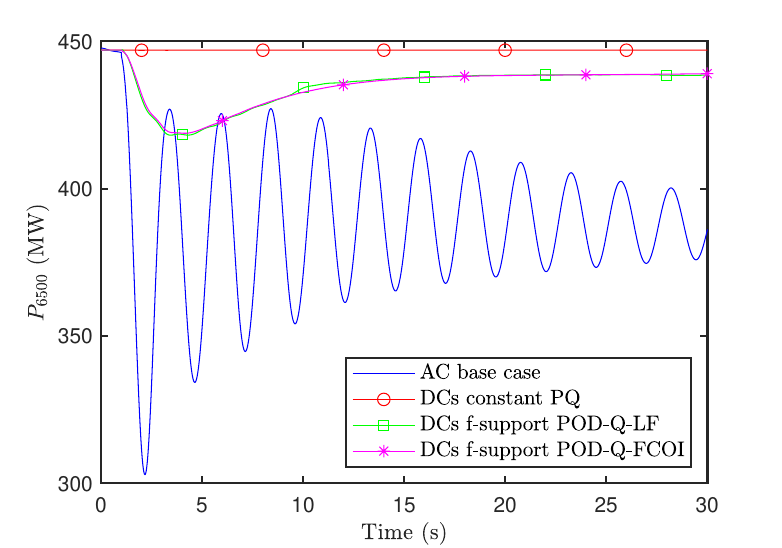}
         \caption{Link A (5100-6500)}
         \label{fig:loss_gene_R1_P6500}
     \end{subfigure}
     \hfill
     \begin{subfigure}[b]{0.37\textwidth}
         \centering
         \includegraphics[width=\textwidth,trim={0cm 0.1cm 0.7cm 0.4cm},clip]{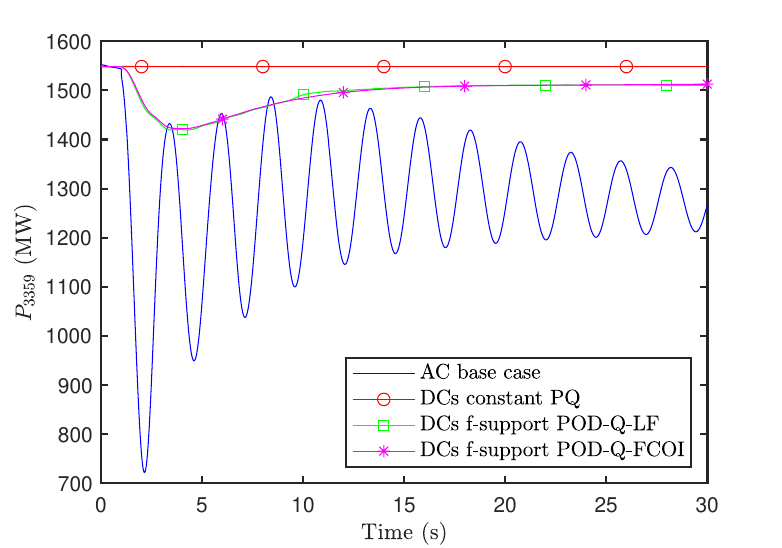}
         \caption{Link B (5101-3359)}
         \label{fig:loss_gene_R1_P3359}
     \end{subfigure}
     \hfill
        \caption{Active power exchange from region 1 to region 2 after the disconnection of a generator at bus 5300 (Region 1).}
        \label{fig:loss_gene_R1_P}
\end{figure}

Fig.~\ref{fig:loss_gene_R1_G} shows the frequency of the two representative generators used above after the disconnection of the generator in Region 1, while Fig.~\ref{fig:loss_gene_R1_P} shows the active power exchange between the two regions (from Region 1 to Region 2). In the AC base case, the frequency nadir and the final frequency are lower than the ones obtained in Region 1, because the two areas are connected synchronously through AC branches and the total frequency-support capability is greater. Naturally, in the AC base case the frequencies at regions 1 and 2 are in synchronism. In the DC segmented case with constant power control, the frequency nadir is low. This is because Region 1 only has frequency support from its synchronous generators and the VSC-HVDC links do not provide frequency support. Notice also that in the case of DC segmentation with constant power the frequency at Region 2 remains constant, because it is decoupled from Region 1 by the VSC-HVDC links. On the contrary in the DC-segmented cases with supplementary controllers, FC in the VSC-HVDC links produces frequency support among the two regions. This causes the frequency nadir to be higher than the one obtained in the DC-segmented case with constant power. The effects of frequency support through the VSC-HVDC links can be observed in Fig.~\ref{fig:loss_gene_R1_P}, where they reduce the power exchange between Region 1 and 2. As a consequence, in the DC-segmented cases with supplementary controllers, the frequency at Region 2 also changes, due to the effect of FC controllers in the VSC-HVDC links. 

Table~\ref{tab:fnadir} shows the frequency nadir ($f_{min}$) and final steady-state frequency ($f_{final}$) after the disconnection of generator G5300-1 in Region 1. Note that the results of the table may not correspond exactly to the ones of Fig.~\ref{fig:loss_gene_R1_G} for the final frequency since the steady state has not been reached in the figure. Results for the disconnection of a generator in Region 2 are also included in the table (G3300-1), which has a nominal power of 1100 MVA and it was injecting 757 MW before the event. As a conclusion, in the DC-segmented cases with supplementary controllers, overall frequency stability is improved, in comparison with the DC-segmented case with constant power.

Additionally, DC segmentation with and without frequency support significantly reduces the change of power flow in the inter area link (see Fig.~\ref{fig:loss_gene_R1_P}). This could avoid tripping of the line in cases where the link is close to its maximum capacity, this could be of interest in cases in where this aspect is critical.

\begin{table}[htb!]
\centering
\caption{Frequency Nadir and steady state after a disconnection of generator in Region 1 or 2 in the four cases.}
\label{tab:fnadir}
\scalebox{0.65}{
\begin{tabular}{@{}c|c|c|c|c|c@{}}
\toprule
Disconnection 	&	Frequency (Hz)	&	AC case	&	DCs constant 	&	DCs f-support 	&	DCs f-support 	\\
of generator in:	&		&		&	PQ	&	POD-Q-LF	&	POD-Q-FCOI	\\ \midrule
Region 1	&	$f_{min}$ (Hz)	&	49.85	&	49.68	&	49.74	&	49.75	\\
	&	$f_{final}$ (Hz)	&	49.97	&	49.94	&	49.94	&	49.94	\\ \midrule
Region 2	&	$f_{min}$ (Hz)	&	49.86	&	49.76	&	49.79	&	49.79	\\
	&	$f_{final}$ (Hz)	&	49.97	&	49.93	&	49.94	&	49.94	\\
\bottomrule
\end{tabular}%
}
\end{table}

\subsection{Impact of communication latency}\label{sec.results.delays}
\noindent The implementation of the POD-Q-FCOI requires a communication system between the generators and the VSCs and its impact on the performance of the controller is analysed here. 

A central control scheme is assumed, where the input error signal of POD-Q-FCOI of Fig.~\ref{fig:PODQ}  with a communication delay reads: 
\begin{equation}\label{eq:PODQ_delay}
	u_{i}=e^{-s \tau} (\omega_{COI,i}-\omega_{i})
\end{equation}
where $\tau$ is the total communication delay.

The delay of (\ref{eq:PODQ_delay}) has been implemented using a first-order Pad\'e's approximation. Realistic values of communication latency delays are analysed ($\tau= 50$ ms and $\tau= 100$ ms).

Table~\ref{tab:eigv_delays} shows the main electromechanical modes of the system for DCs f-support POD-Q-FCOI for different values of communication latency delays. The modes considered are the same as the ones of Table~\ref{tab:eigv}. Results show that the communication delays slightly reduce the damping ratio of the target modes (from 12.40\% to 9.95\% for mode 2 and from 15.40\% to 14.37\% for a delay $\tau = 0.1s$) without significant impact on other modes. Therefore, results prove that POD-Q-FCOI controller is robust against communication latency. Independently of this, as previously discussed, since POD-Q-LF controller uses local signals and both controllers produce similar results, POD-Q-LF controller appears to be a best alternative than POD-Q-FCOI controller, for the test system used in this work.

\begin{table}[htb!]
\centering
\caption{Poorly-damped electromechanical modes of the DCs f-support, POD-Q-FCOI with various communication delays.}
\label{tab:eigv_delays}
\scalebox{0.65}{
\begin{tabular}{@{}c|cc|cc|cc|c@{}}
\toprule
    &   \multicolumn{2}{c}{No delay} &  \multicolumn{2}{c}{$\tau$ = 0.05s} &   \multicolumn{2}{c}{$\tau$ = 0.1s}& Region of \\
N0. &   $\zeta$ (\%) & Freq. (Hz) &   $\zeta$ (\%) & Freq. (Hz)&  $\zeta$ (\%) & Freq. (Hz) & the mode \\ \midrule
2	&	12.40	&	0.83	&	11.14	&	0.84	&	9.95	&	0.84	&	R1	\\
3	&	12.33	&	0.88	&	12.19	&	0.88	&	12.04	&	0.88	&	R2	\\
4	&	15.40	&	0.74	&	14.86	&	0.74	&	14.37	&	0.75	&	R2	\\
5	&	12.16	&	1.07	&	12.19	&	1.07	&	12.22	&	1.07	&	R2	\\
6	&	12.48	&	1.10	&	12.65	&	1.10	&	12.79	&	1.10	&	R1	\\
7	&	16.45	&	0.47	&	17.41	&	0.48	&	18.28	&	0.48	&	R2	\\
8	&	15.61	&	0.97	&	15.10	&	0.98	&	14.11	&	0.99	&	R2	\\
9	&	20.68	&	1.31	&	20.68	&	1.31	&	20.68	&	1.31	&	R1	\\
10	&	14.55	&	1.20	&	13.84	&	1.20	&	13.18	&	1.19	&	R1	\\
\bottomrule
\end{tabular}%
}
\end{table}

\section{Conclusions}\label{sec:conclusions}
\noindent The conclusions obtained in this work can be summarised as follows:
\begin{itemize}
    \item DC segmentation can be very effective to suppress critical inter-area oscillation in stressed AC power systems. However, if VSC-HVDC systems are controlled with constant power: (a) overall frequency stability will be jeopardised, because each asynchronous AC cluster will have lower amounts of frequency support and inertia and (b) intra-area electromechanical oscillations in the AC clusters may have similar damping ratio to those obtained in the AC base case, with room for further improvement.  
    \item Frequency controllers (FC) in the VSC-HVDC links of a DC-segmented system allows frequency support among the different AC clusters, improving overall frequency stability, in comparison with DC segmentation with constant power control. 
    \item Power oscillation damping controllers for the reactive-power injections (POD-Q) in the VSC-HVDC links of a DC-segmented system can be used to damp intra-area oscillations of the asynchronous AC clusters, in addition to the mitigation of the critical inter-area oscillation due to DC segmentation. Two effective POD-Q controllers were analysed and compared: POD-Q-LF (using local measurements), based on previous proposals and POD-Q-FCOI (using global measurements), proposed in this work.
    \item POD-Q-LF (local) and POD-Q-FCOI (global) controllers produced comparable results, and the latter proved to be robust when subject to communication latencies. Nevertheless, taking into account that implementation of POD controllers using remote signals is more complex and expensive than POD controllers using local signals, POD-Q-LF controller seems to be a more practical solution than POD-Q-FCOI controller, for the test system considered.
    \item The concept of numerical eigenvalue sensitivity is proposed and used for the design of POD-Q controllers in the DC-segmented power system, which is an approximation of the theoretical eigenvalue sensitivity and can facilitate its implementation in small-signal-stability tools with limited information of the state matrices. Hence, it has a practical application that could be used for the design of POD controllers in general, and not only in DC-segmented power systems.
    \item The use of DC segmentation, frequency support and POD-Q control brought an important improvement of the overall stability of the test system compared to the initial AC case.
\end{itemize}

\vspace{-0.5cm}
\section*{Appendix}\label{sec:Appendix}

\subsection{AC base case}\label{sec:Appendix-N44_AC_base_case} 
\noindent Dynamic and static data of the Nordic 44 test system can be found under \cite{Nordic44NordpoolN44_BCDyr2022}. The initial power flow condition used in this paper correspond to the Nord Pool data of Tuesday November 10 at 11:38 available at the same link.

\subsection{DC-segmented cases}\label{sec:Appendix-N44_DCs}
\noindent The DC-segmented cases were obtained from the initial Nordic 44 system by replacing AC line 5100-6500 (link B) and the two parallel lines between buses 3359 and 5101 (link A) with VSC-HVDC links. The characteristics of the four VSCs and the two DC lines are included in Table \ref{tab:VSCdataN44}.

\begin{table}[htb!]
\centering
\caption{Data of the HVDC links in the DC-segmented Nordic 44 system.}
\label{tab:VSCdataN44}
\scalebox{0.65}{
\begin{tabular}{p{6cm}p{4cm}}
\toprule
Parameters 	\\		
(p.u's: converter rating) 	&	Values	\\ \midrule
Rating VSC (VSCA1-2/ VSCB1-2)	&	3500/800 MVA	\\
DC voltage (VSCA1-2/ VSCB1-2)	&	$\pm$ 535/320 kV	\\
Configuration	&	Symmetrical monopole	\\
Max active power (VSCA1-2/ VSCB1-2)	&	$\pm 3500/800 MW$	\\
Max reactive power (VSCA1-2/ VSCB1-2)	&	$\pm 1400/320 MVAr$	\\
Max. current	&	1 p.u.	\\
Max. DC voltage $V^{max}_{dc,i}, V^{min}_{dc,i}$	&	1.1, 0.9 p.u.	\\
Current-controller time constant ($\tau$)	&	2 ms	\\
Connection imp. ($R_{s,i} + jX_{s,i}$)	&	0.004 + j 0.2 p.u.	\\
Outer control gains	&	\\	
P control	& 	$i_{ac,d,i}^{ref}=P_{ac,i}^{*}/V_{ac,i}$ \\
V$_{dc}$ prop./int.($K_{p,d2}/K_{i,d2}$)	&	10 p.u./20 p.u./s	\\
Q control	&	$i_{ac,q,i}^{ref}=-Q_{ac,i}^{*}/V_{ac,i}$ \\
VSCs' loss coefficients	&	a = b = c = 0 p.u.	\\
DC-bus capacitance ($C_{dc,i}$) (VSCA1-2/ VSCB1-2)	&	305/195 $\mu F$	\\
DC-line series resistance, inductance ($R_{dc,ij}, L_{dc,ij}$)	&	\\	
 - link A (5101-3359)	&	1.6 $\Omega$, 67 mH	\\
 - line B (5100-6500)	&	7.2 $\Omega$, 258 mH	\\
 \\
\end{tabular}%
}
\end{table}

VSCs A2 (3359) and B2 (6500) were set in $V_{dc}$-control mode while VSCs A1 (5101) and B1 (5100) were set in $P$-control mode.

Parameters of Frequency Controller (FC) (Fig.~\ref{fig:fsupport} of Section~\ref{sec.FC}) are $K_{FC,i}=100$ pu, $T_{FC,i}=0.1$ s and $\pm \Delta P_{ac,i}^{max}= \pm 1$ pu.

\section*{Acknowledgments}
\noindent The work of Mr. Mathieu Robin is within a collaboration of SuperGrid institute in the doctoral programme of Comillas Pontifical University.

\bibliographystyle{IEEEtran}
\bibliography{articlePOD.bib}

\textbf{Mathieu Robin} obtained the M.Eng degree from Ecole Centrale Lyon, Ecully, France  and  the  M.Sc  degree  in Electrical Engineering from jointly Ecole Centrale Lyon and Claude Bernard University in Lyon, France, both in 2019. He is actually pursuing his Ph.D. degree at Universidad Pontificia Comillas de Madrid (UPCO), Madrid, Spain and is working as a research engineer at SuperGrid Institute, Villeurbanne, France. 

His research interests are stability, modelling and control of power systems, particularly VSC-HVDC systems and DC segmentation.

\textbf{Javier Renedo} (Senior Member, IEEE) received his M.Sc. degree in electrical engineering from Universidad Pontificia Comillas de Madrid (Comillas, for short), Spain, in 2010, the M.Sc. degree in mathematical engineering from Universidad Carlos III de Madrid, Spain, in 2013, and the Ph.D. degree in modelling of engineering systems from Comillas, in 2018. His research interests include power system stability, VSC-HVDC systems, and power systems with large amounts of renewable resources.

\textbf{Juan Carlos Gonzalez Torres} received both the Electromechanical Engineering degree from Universidad Autonoma de San Luis Potosi and the Master of Engineering from Ecole Centrale de Lille in 2014. He graduated with a Master of Science degree from ParisTech  (The Paris Institute of Technology)  in 2015. In 2019, he obtained the Ph.D. in Automatic Control from Paris-Saclay University.  He has been working as a research engineer at Supergrid Institute since 2016,  where  he is part of the Architecture \& Systems program.  His main research interests include modeling and control of power systems, HVDC transmission systems and integration of renewable energies via power electronics-based devices.

\textbf{Aurelio Garcia Cerrada} (Senior Member, IEEE from 2015) M.Sc. (1986) from the Universidad Politécnica de Madrid, Spain, and Ph.D. (1991) from the University of Birmingham, U.K. He is a Professor in the Electronics, Control Engineering and Communications Department and a member of the Institute for Research in Technology (IIT) at the Universidad Pontiﬁcia Comillas de Madrid. His research focuses on power electronics and its applications to electric energy systems.

\textbf{Luis Rouco} (Senior Member, IEEE) received the “Ingeniero Industrial and “Doctor Ingeniero Industrial degrees from Universidad Politécnica de Madrid in 1985 and 1990 respectively. He is Professor of electrical engineering with the Department of Electrical Engineering of the School of Engineering of Universidad Pontificia Comillas. He served as Head of the Department from 1999 through 2005. He develops his research activities at Instituto de Investigación Tecnológica (IIT). Prof. Rouco is a Distinguished Member of Cigré and Member of the Executive Committee of Spanish National Committee of Cigré. He has been Visiting Scientist at Ontario Hydro, MIT and ABB Power Systems.

\textbf{Abdelkrim Benchaib} received his Ph.D. from the Automatic Systems Laboratory, Picardie University, France, in December 1998 and in 2014 his HDR (French postdoctoral degree allowing its holder to supervise PhD students) from Paris-Orsay University.  He joined Alstom in July 2000 where he has been working as a power quality and smart grid project leader and thereafter with GE Grid solutions. Currently, he is seconded by GE to work with the SuperGrid institute where he is a Sub-program group leader for real-time strategies of super grids AC/DC control and dispatch. His expertise and research interests include automatic control, AC and DC power systems and power electronics. Dr. Benchaib is associate Professor at the Cnam (Conservatoire National des Arts et Metiers) teaching wind energy and power network. He has been General Chairman of the EPE ECCE Europe Conference for its 2020 edition (EPE conference is one of the bigest events in the world for Power Electronics with up to 1000 participants) . Abdelkrim Benchaib is the secretary of the EPE association.

\textbf{Pablo Garcia Gonzalez} holds a M.Sc. in electrical engineering (1992) and a PhD (2000), both from the Universidad Pontificia Comillas, Madrid, Spain. He is a Full Professor at the ICAI School of Engineering. His teaching and research activities focus on applications of power electronics, control systems, and the integration of distributed energy resources in power systems.




\end{document}